\documentclass[12pt]{article}
\usepackage{hyperref, graphicx, color, amssymb, amsmath, cite}
\usepackage[margin=1.0in]{geometry}

\newcounter{si}

\begin{document}
\title{Are slot and sub-wavelength grating waveguides better than strip waveguides for sensing?}
\author{
        Derek Kita, J\'{e}r\^{o}me Michon, Juejun Hu \\
        Department of Materials Science \& Engineering\\
        Materials Research Laboratory\\
        Massachusetts Institute of Technology\\
        Massachusetts Ave, Cambridge MA 02139, USA
            \and
        Steven G. Johnson\\
        Department of Mathematics\\
        Massachusetts Institute of Technology\\
        Massachusetts Ave, Cambridge MA 02139, USA
}
\date{\today}


\maketitle

\begin{abstract}
The unique ability of slot and sub-wavelength grating (SWG) waveguides to confine light outside of the waveguide core material has attracted significant interest in their application to chemical and biological sensing. However, high sensitivity to sidewall roughness induced scattering loss in these structures compared to strip waveguides casts doubt on their efficacy. In this article, we seek to settle the controversy for silicon-on-insulator (SOI) photonic devices by quantitatively comparing the sensing performance of various waveguide geometries through figures of merit that we derive for each mode of sensing. These methods (which may be readily applied to other material systems) take into account both modal confinement and roughness scattering loss, the latter of which is computed using a volume-current (Green's-function) method with a first Born approximation. For devices based on the standard 220\,nm SOI platform at telecommunication wavelengths ($\lambda=1550$\,nm) whose propagation loss is predominantly limited by random line-edge sidewall roughness scattering, our model predicts that properly engineered TM-polarized strip waveguides claim the best performance for refractometry and absorption spectroscopy, while optimized slot waveguides demonstrate $>5\times$ performance enhancement over the other waveguide geometries for waveguide-enhanced Raman spectroscopy.\end{abstract}

\section{Introduction}
Waveguide based chemical and biological sensing is rapidly advancing as a prime application area for integrated photonics. However, there is currently no universally agreed-upon waveguide geometry that optimally enhances detected signals in the presence of fabrication-induced scattering losses.  In classical strip waveguide sensors (Fig.~\ref{fig:3d_waveguides}), molecules of interest interact with the relatively weak evanescent electric field outside the waveguide core. To boost light--molecule interactions, slot waveguides~\cite{Barrios2007,Claes2009,Carlborg2010,DellOlio2007,Liu2013,Barrios2008} and SWG waveguides~\cite{Wanguemert-perez2014,Flueckiger2016,Yan2016} have been proposed and demonstrated as alternative sensing platforms. In these waveguides, a larger fraction of the mode resides in the low-index cladding (usually the sensing medium, such as air or water) where the molecules are located. Indeed, improved refractive-index sensitivity (defined as the induced wavelength detuning per unit refractive-index change in the surrounding media) has been experimentally validated in resonator refractometry sensors based on both slot and SWG waveguides~\cite{Barrios2007,Claes2009,Flueckiger2016}, and enhanced Raman conversion efficiency per unit device length has also been measured in slot waveguides~\cite{Dhakal2015}. Nevertheless, the benefits may be offset by their increased susceptibility to optical loss induced by sidewall-roughness scattering.  Many authors have studied roughness loss in dielectric waveguides both theoretically (through coupled-mode theory and the volume-current / Green's-function methods~\cite{Marcuse1969a, Snyder1983, Payne1994, Johnson2005}) and experimentally~\cite{Lee2000, Lee2001, Morichetti2010, Lee2015}, but these analyses focused only on minimizing the power scattered into the far field.  This tradeoff between mode confinement and attenuation is typically reported for plasmonic waveguide structures~\cite{Berini2006, Barnes2006, Offermans2011, Zafar2015, Islam2017, Zhang2018} and some resonant sensor devices~\cite{Hu2009a, Nitkowski2011a, Huang2014, Subramanian2015}, but differences in fabrication conditions often make it difficult to perform side-by-side comparisons.  Efficiently quantifying the tradeoff between mode delocalization and scattering loss is critically important for the rational design of chip-scale photonic sensors (as well as electro-optic modulators~\cite{Koos2009} and light sources~\cite{Guo2012}) and to date there seems to be no work that quantifies these two effects with a single figure of merit (FOM) for strip, slot, and SWG waveguides.

In this paper, we develop easily computed figures of merit (Sec.~\ref{sec:sensingFOM}) that capture the precise trade-off between field confinement and roughness-induced scattering loss for waveguide-based sensing applications. The key to efficiently evaluating our figures of merit is perturbation theory~\cite{Snyder1983, Johnson2002}, both to evaluate the impact of mode confinement on sensing (see Sec.~\ref{sec:sensingFOM} and \href{link}{Supplemental Information (SI)}) and also to evaluate roughness scattering by computing the power radiated by equivalent current sources along the surface~\cite{Johnson2005, Prigogine1994} (Sec.~\ref{sec:scattering}). The latter approach circumvents costly and roughness-dependent direct simulation of disordered waveguides, giving us a figure of merit comparison independent of the precise roughness statistics as long as the roughness correlation length is sub-wavelength. Although computation of the ``effective'' sources to represent roughness scattering is in general quite complicated~\cite{Johnson2005}, we both propose a powerful general approach and we demonstrate that, for typical experimental roughness, a simple ``locally flat perturbation'' approximation is accurate (Sec.~\ref{sec:scattering} and \href{link}{SI}). The results of this work indicate that larger modal overlap with the sensing medium does not always correspond to better performance (due to increased scattering losses). By evaluating our figures of merit for strip, slot, and sub-wavelength grating (SWG) designs for bulk/surface/Raman sensing (Sec.~\ref{sec:comparison}), we obtain the results that for typical SOI waveguides at a wavelength of $\lambda$=1550\,nm (1) the simple TM-polarized strip waveguide is $>3\times$ better than other geometries considered here for bulk absorption sensing and refractometry, (2) the TM-polarized strip waveguide and TE-polarized slot waveguide are both $>3\times$ better than other geometries for surface-sensitive refractometry and absorption sensing, and (3) the TE-polarized slot waveguide is $>5\times$ better than other geometries for Raman sensing.  Furthermore, we provide an extensive comparison (Sec.~\ref{sec:experimental-comparison}) to published experimental results, in which we applied our numerical method to compute the ratio of losses for pairs of waveguide geometries that were fabricated and reported in the literature, and our predictions exhibit good agreement to within experimental accuracy. As discussed in our concluding remarks (Sec.~\ref{sec:conclusion}), we believe that these comparisons and figures of merit will drive future experimental and theoretical exploration of new waveguide geometries for the purpose of sensing.

\section{Sensing Figures of Merit}\label{sec:sensingFOM}

Here we consider three sensing techniques: 1) surface-sensitive refractometry and absorption sensing, where a sensor surface is coated with a binding agent (e.g. antibody) that specifically attaches to the target molecule and the device monitors the change in index or optical absorption caused by monolayer or few-layer molecular binding on the surface; 2) bulk index/absorption sensing, where the sensor detects the change in bulk index/absorption in the adjacent sensing medium; and 3) waveguide-enhanced Raman spectroscopy (WERS)~\cite{Dhakal2014,Holmstrom2016,Evans2016}, where the excitation light and spontaneous Raman emission signal from molecules in the surroundings co-propagate in a waveguide. In all cases, the sensor performance is related to the external confinement factor and divided by the attenuation coefficient.  This ``benefit-to-cost'' ratio is described in~\cite{Subramanian2015} for absorption spectroscopy in silicon and silicon nitride waveguides and in~\cite{Berini2006} for plasmonic waveguides, where there is a similar drive to maximize waveguide confinement and minimize the strong attenuation coefficients of surface plasmon polaritons~\cite{Barnes2006}.  Equivalently, for optical resonator refractometers the appropriate figure of merit is often stated as the ratio of the refractive-index sensitivity to the resonance full-width at half maximum (FWHM) (which is proportional to the attenuation coefficient)~\cite{Hu2009a, Nitkowski2011a, Offermans2011, Huang2014, Zafar2015, Islam2017, Zhang2018}.  For most prior literature that analyzes sensing in non-resonant waveguide devices (and also some resonant devices~\cite{Hanumegowda2005, Claes2009, Passaro2012}), the quantity of interest typically reported is the fractional power change for a small increase in analyte (i.e. the numerator of our FOM)~\cite{Heideman1990, Lu2012, Liu2013, Singh2014a, Kozma2014, Dullo2015, Bastos2018}.  This is likely due to the difficulty in precisely measuring the attenuation coefficient or scattering loss for a single evanescent waveguide sensor or Mach--Zehnder refractometer~\cite{Lee2000, Bojko2011} (in contrast, for resonant devices this information is immediately available in the resonance FWHM)~\cite{Ryckeboer2014}.

\begin{figure*}[ht!]
\centering
\fbox{\includegraphics[width=0.95\textwidth]{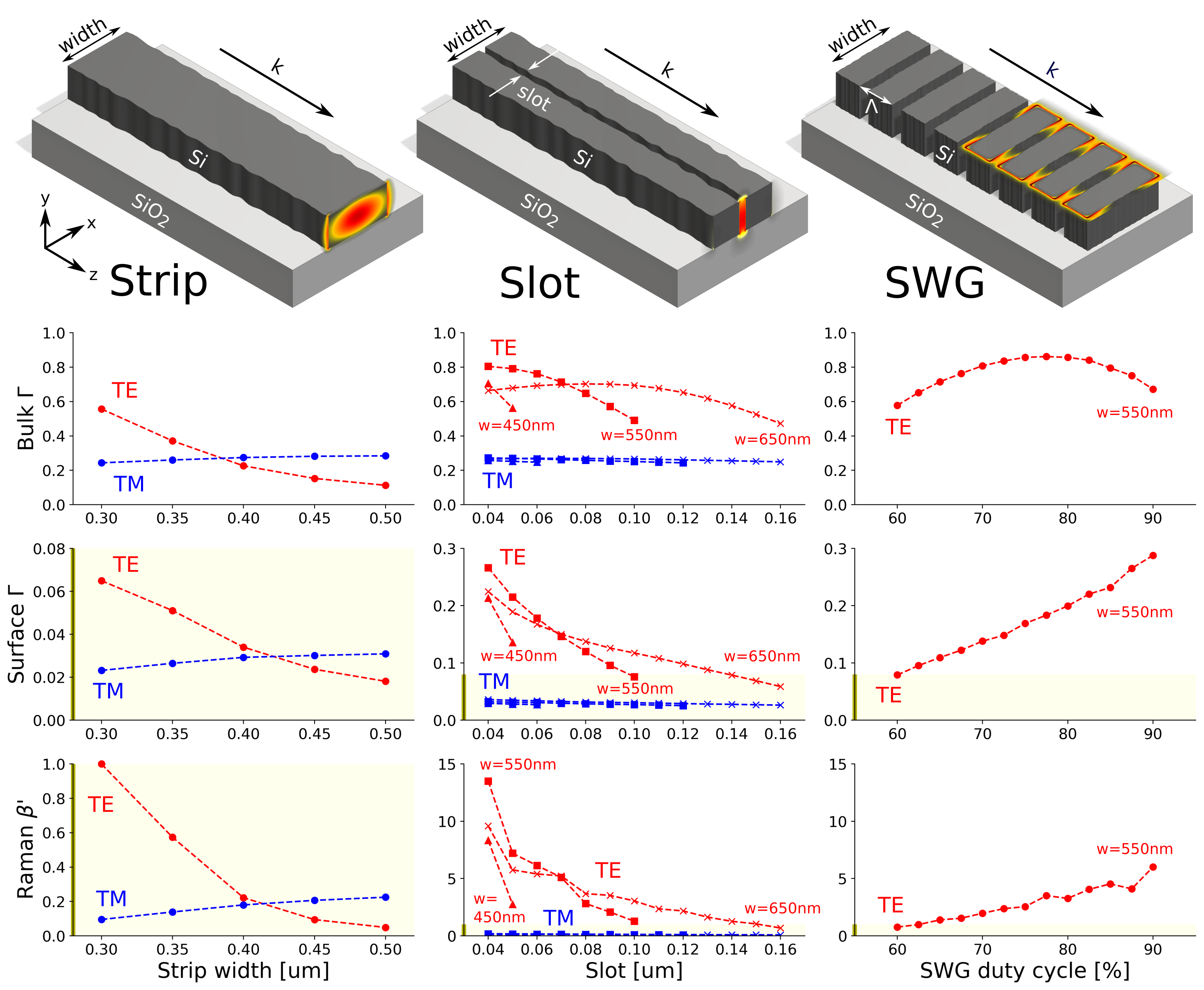}}
\caption{Top row: vertically symmetric surface roughness along a silicon strip waveguide on silicon dioxide substrate, a slot waveguide, and a SWG waveguide.  The electric field magnitude $|\vec{E}|^2$ is overlayed at several cross-sections.  Second (from top) to fourth rows: external (bulk) confinement factor $\Gamma$, surface confinement factor $\Gamma$ (field integral over 8\,nm thin region at air/solid interface), and the \emph{normalized} Raman gain coefficient $\beta'$, respectively, for each single-mode waveguide geometry as a function of the relevant design parameters (total width, slot size, and duty cycle).  Shaded regions denote the relative y-axis scaling between adjacent plots in each row.  For slot waveguides, the total width is denoted as follows: $\triangle$=450\,nm, $\square$=550\,nm, and $\times$=650\,nm.  For SWG waveguides, the width and period are fixed to 550\,nm and 250\,nm, respectively. The surface $\Gamma$ and bulk $\Gamma$ are calculated via first-order perturbation theory (which is exactly equivalent to Eq.~\ref{eq:gamma}) with a resolution of 256\,pixels/$\mu$m and for SWG waveguides with a resolution of 128\,pixels/$\mu$m.  The Raman gain coefficients are computed via Eq.~\ref{eq:beta} and~\ref{eq:betaSWG} in 3-dimensions with a resolution of 128\,pixels/$\mu$m.}\label{fig:3d_waveguides}
\end{figure*}

The general goal in the first two sensing modes is to maximize the device sensitivity, which is the change in fractional optical power ($\Delta P / P_\text{input}$) induced by a small change in the number of analyte molecules that alter the cladding absorption coefficient or index of refraction.  Thus, the relevant metric is the sensitivity in units of inverse number density.  For example, as described in the \href{link}{SI}, the absorption sensitivity is proportional to $\Gamma_\text{clad}\alpha_0/\alpha_s$, where $\Gamma_\text{clad}$ is either the external confinement factor of the entire cladding region for bulk sensing or the surface region (approximated by a thin volume next to the interface), $\alpha_s$ is the scattering loss per unit length, and $\alpha_0$ is the molecular absorption coefficient (related to the bulk absorption coefficient via $\alpha_\text{abs} = m \alpha_0$, where $m$ is the number density of analyte molecules).  In general, we find that the only geometry dependence occurs in a factor $\Gamma_\text{clad}/\alpha_s$, leading us to the dimensionless figure of merit for surface sensitive and bulk absorption/index sensing:
\begin{equation}
\text{FOM}_\Gamma = \frac{\Gamma_\text{clad}}{\alpha_s\lambda} \label{eq:FOMabs}
\end{equation} 
where $\lambda$ is the vacuum wavelength of light.  In the third sensing mode (Raman sensing), the goal is to maximize the power of Raman scattered light collected at the output of a waveguide for a given input laser power.  As such, the corresponding figure of merit is the dimensionless quantity derived in the \href{link}{SI}:
\begin{equation}
\text{FOM}_\beta = \frac{\beta}{\alpha_s} \label{eq:FOMram}
\end{equation}
where $\beta$ is the Raman gain coefficient (with units of 1/length) defined as the power of evanescently scattered Raman light collected back into the waveguide per unit length and normalized by the input power. The external confinement factor~\cite{Weber1974, Kumar1981, Mittra1986, She1990, Gupta1992, Themistos1995, Johnson2001a, gamma} as well as the Raman gain coefficient are quantified via perturbation theory (see \href{link}{SI}):
\begin{align}
\Gamma_\text{clad} &= \frac{dn_\text{eff}}{dn_\text{clad}} = \frac{n_g}{n_\text{clad}}\frac{\int_\text{clad} \epsilon |\vec{E}|^2 d^2x}{\int \epsilon |\vec{E}|^2 d^2x}\label{eq:gamma}\\
\beta &= \frac{\alpha_\text{ram}^2 \omega_n^2 m}{4c^2} \cdot \frac{n_g^2(\omega)}{n_\text{clad}^2} \frac{\int_\text{clad}|\vec{E}(x, \omega)|^4 d^2x}{(\int \epsilon(x)|\vec{E}(x,\omega)|^2 d^2x)^2} \label{eq:beta}
\end{align}
where $n_\text{clad}$, $n_\text{eff}$, and $n_g$ are the cladding material, effective, and waveguide group indices, $\epsilon$ is the permittivity, $E$ is the electric field of the waveguide mode, $\omega$ is the input angular frequency, integration is over the cross section of the waveguide, restricted in the numerator to the cladding/sensing region, and $c$ is the speed of light.  We also assume that the Raman shift is relatively small so that $|\Delta \omega| \ll \omega$.  For periodic structures such as SWG waveguides, integration is performed over the volume of a unit cell with period $\Lambda$:
\begin{align}
\beta_\text{SWG} &= \frac{\alpha_\text{ram}^2 \omega_n^2 m}{4c^2} \cdot \frac{n_g^2(\omega)}{n_\text{clad}^2} \frac{\Lambda \cdot \int_\text{clad}|\vec{E}(x, \omega)|^4 d^3x}{(\int \epsilon(x)|\vec{E}(x,\omega)|^2 d^3x)^2} \label{eq:betaSWG}
\end{align}
We note here that the optimized length of the waveguide for $\text{FOM}_\Gamma$ is $z=1/(\alpha_s + \Gamma_\text{clad}\alpha_\text{abs}) \sim 1/\alpha_s$ and for $\text{FOM}_\beta$ is $z=1/\alpha_s$. In this work we assume that sidewall roughness scattering constitutes the dominant source of optical loss in waveguides, which is typically the case in high-index-contrast waveguide systems at moderate optical powers such as SOI~\cite{Melati2014, Borselli2005, Lee2012}, silicon nitride on insulator~\cite{Bauters2011a, Worhoff2015}, chalcogenide glass on oxide~\cite{Hu2007}, and TiO$_2$ on oxide~\cite{Evans2015}. We also note that each FOM is generic to many different sensing device configurations, such as serpentine/spiral waveguides, ring resonators, and Mach--Zehnder interferometers.

In all three cases, the sensor limit of detection (LOD) is determined by the modal overlap with the sensing region as well as the optical path length, the latter of which is limited by the waveguide propagation loss. The surface or bulk modal confinement factors and the Raman gain coefficient are readily computed from standard frequency-domain eigenmode solvers and the results are plotted in Fig.~\ref{fig:3d_waveguides}, which clearly show that slot and SWG structures indeed significantly enhance modal overlap with the sensing medium.

\section{Scattering-loss calculations}\label{sec:scattering}
In the volume-current method, waveguide perturbations can be described to
first order (neglecting multiple-scattering effects) by dipole moments (polarizations)
induced in the perturbation by the original (unperturbed) waveguide field $\vec{E}_0$.
These perturbations act as current sources that create the scattered field. For example, a small perturbation $\Delta \epsilon$ in the permittivity acts like a current source $\vec{J} = \Delta \epsilon \vec{E}_0$~\cite{Marcuse1969a, Lacey1990}. However, for perturbations in high-index-contrast waveguide interfaces, calculating the induced polarization (and hence the effective current source) is in general much more complicated~\cite{Johnson2005}, and requires solving a quasistatics problem (Poisson's equation)~\cite{Jackson1999}. For a given statistical distribution of the surface roughness, we show in \href{link}{SI} how we can solve a set of quasistatics problems to compute the corresponding statistical distribution of the polarization currents. Fortunately, however, there is a simplification that applies to typical experimental regimes ($L_c > 10 \sigma$, where $L_c$ is the correlation length of the surface roughness and $\sigma$ the root mean square roughness amplitude).  If $L_c$ (typically $50$--$100$\,nm for SOI waveguides) is much larger than the amplitude of the roughness ($0.5$--$2$\,nm in state-of-the-art devices~\cite{Xia2007, Wood2014,Lee2015,Wang2014}), then one can approximate the surface perturbation as locally flat. In this case, there is an analytical formula for the induced current from a locally flat interface shifted by a distance $\Delta h$~\cite{Jackson1999}:
\begin{equation}
\vec{J} = -i\omega \Delta h (\Delta \epsilon E_\parallel - \epsilon \Delta (\epsilon^{-1})D_\perp)\delta(\vec{x})
\label{eq:currentstrength}
\end{equation}
(Note that the currents depend on the orientation of $\vec{E}_0$ relative to the interface: $E_\parallel$ is the component of $E_0$ parallel to the surface and $D_\perp$ is the component of $\epsilon E_0$ perpendicular to the surface.) In this case, the correlation function of the current $J$ is simply proportional to the correlation function of the surface profile.  We verify in \href{link}{SI} that the full quasistatic calculation reproduces this locally flat interface approximation in typical experimental regimes.

Given these currents and their statistical distribution (from the statistics of $h$), one can then perform a set of Maxwell simulations on the unperturbed waveguide geometry (hence, at moderate spatial resolution) to find the corresponding radiated power. Again, there is a somewhat complicated procedure in general to compute the effect of currents with an arbitrary correlation function.  But, in the case where the correlation length ($\sim 100$\,nm) is much less than the wavelength ($\sim 1-2$\,$\mu$m), we can approximate this ``colored'' noise distribution by uncorrelated ``white'' noise~\cite{Prigogine1994}, effectively treating the currents as uncorrelated point sources with the same mean-squared amplitude $\langle J^2\rangle$.  Since the power radiated by uncorrelated point sources is additive, we can then simply average the power radiated from different points on the surface to find the scattering loss per unit length $\alpha_s$~\cite{Johnson2005}.  Finally, the computation simplifies even further because we only compute the ratio of the FOMs for different waveguides, in which case all overall scaling factors and units (including all dependence on the roughness statistics $L_c$ and $\sigma$) cancel out.

\begin{figure*}[tb!]
\centering
\includegraphics[width=\textwidth]{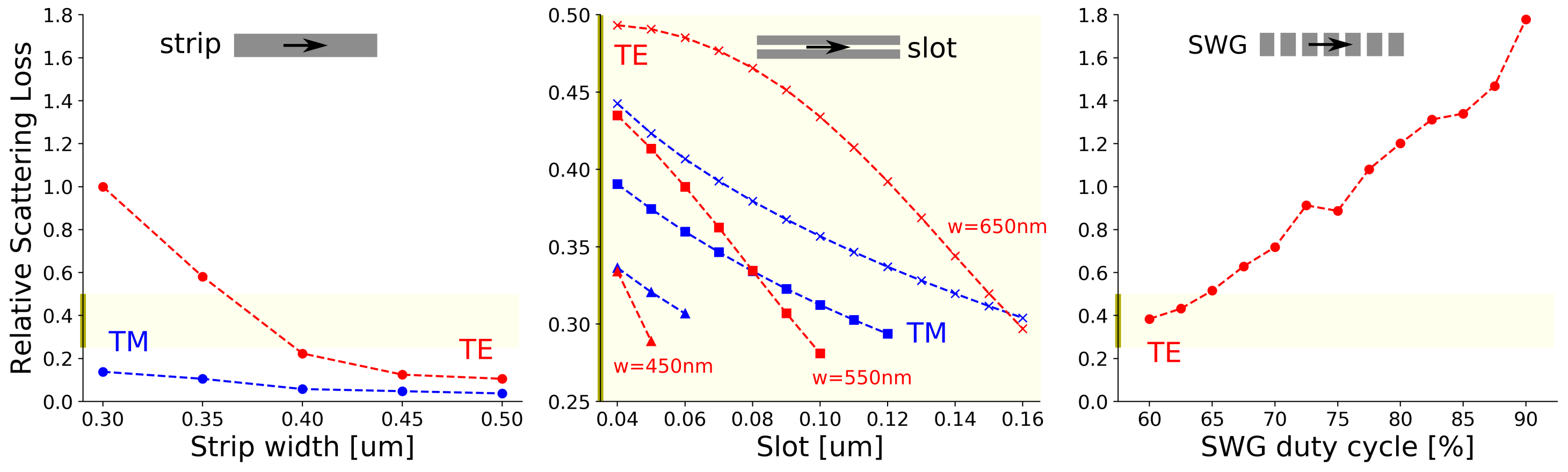}
\caption{Relative scattering loss $\alpha_s / \max(\alpha_{s,\text{TE-strip}})$ computed by FDTD simulations of vertical ($y$-direction) lines of dipole moments and averaged over all sidewall positions.  Dipole moment amplitudes were computed via Eq.~\ref{eq:currentstrength} with incident field strength and phase determined by numerically computed mode profiles.  Results shown for TE (red) and TM (blue) strip (left), slot (middle), and SWG (right) waveguides.  For slot waveguides, the total width is denoted as follows: $\triangle$=450\,nm, $\square$=550\,nm, and $\times$=650\,nm.}\label{fig:ScatteringLoss}
\end{figure*}

\section{Comparison of Waveguide Geometries}\label{sec:comparison}

Using the aforementioned volume-current method, we quantified the relative scattering losses ($\alpha_s$) for a large set of strip, slot, and SWG waveguide geometries for various design parameters such as width, slot size, and duty-cycle.  All waveguide geometries (Fig.~\ref{fig:3d_waveguides}) are assumed to be fabricated from SOI wafers with 220\,nm Si layer thickness, negligible scattering losses at the top and bottom flat surfaces, and for a target wavelength of $1550$\,nm.  The calculations, described in detail in \href{link}{SI}, consist of only two computationally inexpensive steps.  First, we compute the electric ($\vec{E}$) and displacement ($\vec{D}$) field profiles using a frequency-domain eigenmode solver~\cite{Johnson2002}, which is used to determine the confinement factors, Raman gain coefficient, and the strength of the induced dipole moments via Eq.~\ref{eq:currentstrength}.  For each unique sidewall position and for each waveguide geometry, we use a single three-dimensional finite-difference time-domain (FDTD) simulation with a vertical line of dipole moments (corresponding to vertically symmetric sidewall roughness, i.e. line-edge roughness~\cite{Barwicz2003a, Patsis2003, Constantoudis2003}) to compute the scattered power from both far-field radiation and reflection.  Finally, the scattered power is averaged over all inequivalent sidewall positions and computed per unit length along the direction of propagation.  The relative scattering loss per unit length for the considered geometries is shown in Fig.~\ref{fig:ScatteringLoss}.

For strip waveguides, we analyzed the TE-like and TM-like polarized modes for various waveguide widths.  Results for slot waveguides of TE and TM polarization are computed for various total waveguide widths and air-slot gaps.  For SWG waveguides, the width and period are fixed at 550\,nm and 250\,nm respectively, and only TE ($x$-antisymmetric) modes are considered as a function of the duty cycle, which is defined by $\text{DC} = (\Lambda-s)/\Lambda$ where $\Lambda$ is the grating period and $s$ is the size of the air gap along the $z$ direction.  The particular width of 550\,nm and polarization was chosen for this analysis since the corresponding waveguide geometries exhibit single-mode behavior for a relatively large range of duty cycles, as confirmed by numerically computed dispersion diagrams.  All waveguide dimensions are chosen such that no more than one TE and one TM mode exists.

Our results indicate that slot and SWG waveguides do in fact greatly increase scattering losses relative to standard 450\,nm wide TE and TM strip waveguides~\cite{Dumon2004}.  The scattering loss computed by the volume-current method considers both the incident field strength at the location of the perturbation and the local density of states (LDOS)~\cite{Oskooi2013}, which is determined by the surrounding geometry.  (As such, current sources at the outer edges of a strip waveguide will radiate different amounts of power than current sources in the air-gap region of a slot waveguide.)  Our comparison of scattering losses for different geometries, as shown in Fig.~\ref{fig:ScatteringLoss}, is largely consistent with prior experimentally measured values of surface roughness in SOI waveguides~\cite{Sardo2008, Alasaarela2011, Debnath2016a}.  However, a direct comparison between experiment and theoretically/numerically computed values is difficult since experimental uncertainties are typically on the order of a few dB/cm.  In addition, an accurate comparison of scattering losses for different geometries requires side-by-side fabrication of waveguides on the same process and material platform, since different processes introduce dramatically different roughness statistics.  As such, literature for this is relatively scarce, and numerical methods provide a means for quickly evaluating and comparing new waveguide structures.

\begin{figure*}[tb!]
\centering
\includegraphics[width=\textwidth]{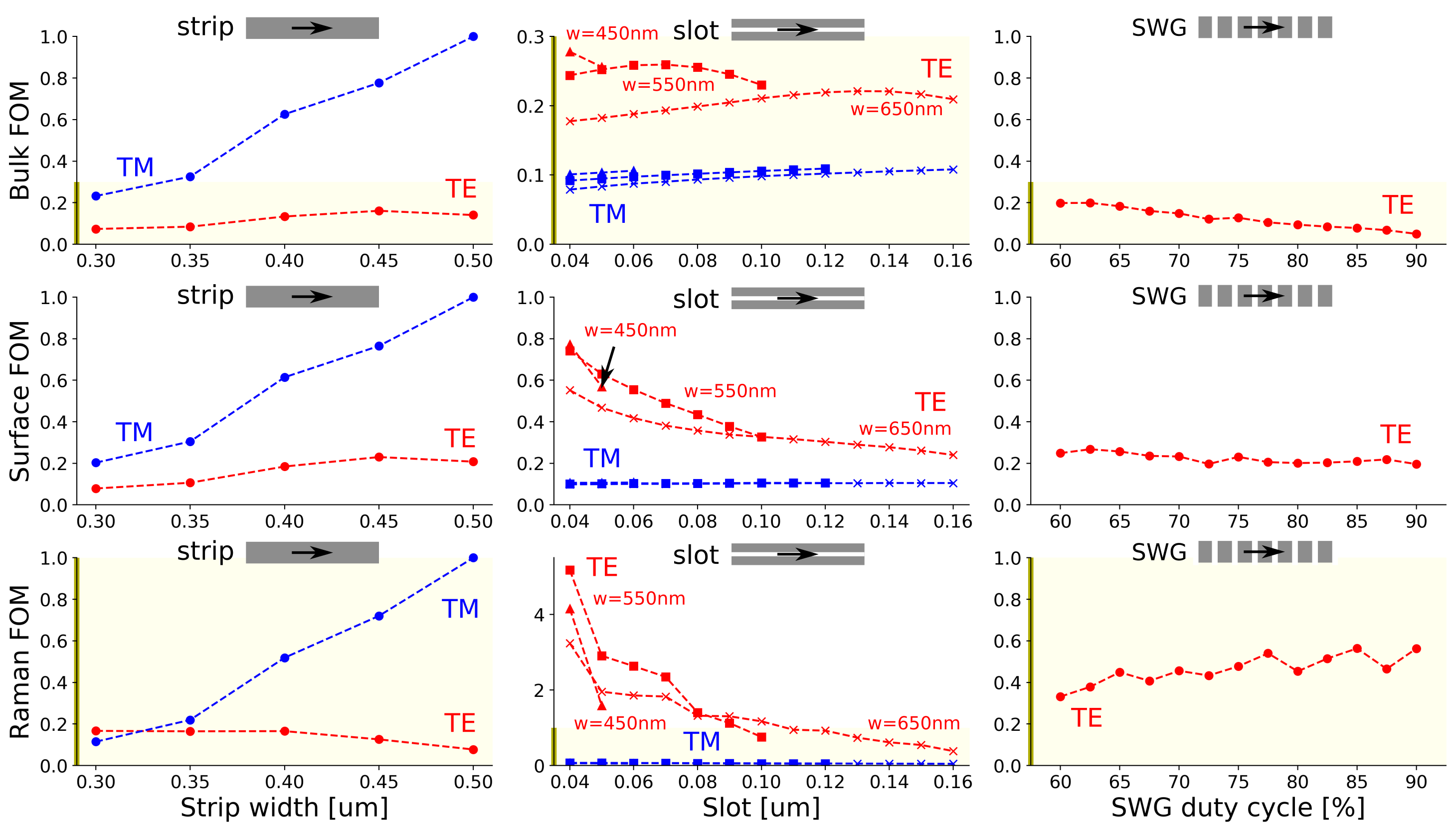}
\caption{Side-by-side performance comparison for strip (left column), slot (middle column), and SWG (right column) waveguides in the presence of slow-varying ($L_c > 10\cdot\sigma$) Gaussian random roughness.  Normalized absorption sensing figure of merit $\text{FOM}_\Gamma$ and normalized Raman gain coefficient $\text{FOM}_{\beta'}$ calculated via Eq.~\ref{eq:FOMabs} and~\ref{eq:FOMram}, respectively, as a function of the relevant design parameters (total width, slot size, and duty cycle).  For slot waveguides, the total width (as depicted in Fig~\ref{fig:3d_waveguides}) is denoted as follows: $\triangle$=450\,nm, $\square$=550\,nm, and $\times$=650\,nm.}\label{fig:FOMresults}
\end{figure*}

With the computed scattering losses, confinement factors, and Raman gain coefficients, we then computed the relevant FOM for each mode of sensing (via Eq.~\ref{eq:FOMabs} and~\ref{eq:FOMram}), which is presented in Fig.~\ref{fig:FOMresults}.  Our results indicate the TE slot and SWG structures do in fact provide modest improvements in sensing performance over traditional TE-polarized strip waveguides.  However, TM-polarized strip waveguides (which are here assumed to have negligible roughness on the top and bottom interfaces) exhibit significantly higher performance owing to their reduced propagation loss and longer accessible optical-path length.  For bulk and surface absorption sensing, the performance of the SWG waveguides gradually decrease with increase in duty cycle (as the air-slot region becomes smaller), indicating that the increase in scattering loss in small SWG gap-regions outweighs the benefits provided by field localization in the air gap.  On the other hand, slot-waveguide structures demonstrate improved performance as the slot size decreases (narrow-slot waveguides show $2\times$ improvement over large-slot waveguides for surface sensing and $5\times$ improvement for Raman sensing), with the exception of bulk absorption sensing in the air region.  For bulk absorption sensing, there appears to be a critical slot size (70\,nm slot for the 550\,nm wide waveguides, and 130\,nm slot for the 650\,nm wide waveguides) below which scattering losses dominate the FOM and above which the confinement factor is suboptimal.

For Raman spectroscopy, the gain coefficient is related to the fourth power of electric field rather than the square of the electric field, so regions of high electric field exhibit significant performance enhancements, as shown by the numerically computed values of the relative Raman gain coefficient $\beta'$ in Fig.~\ref{fig:3d_waveguides}. In our computations, we find that waveguides with narrow slots do in fact tend to produce sufficiently more Raman signal than what is lost due to sidewall scattering at the silicon/air interface.  Wide (550\,nm) silicon waveguides with narrow (40\,nm) air slots yield the highest bulk Raman FOM by a factor of $8 \times$ over SWG waveguides and $5\times$ over TM strip waveguide modes. Despite having an improved Raman gain coefficient, the SWG structures suffer from significantly higher scattering losses due to the increased sidewall surface area.

\section{Comparison with experiment}\label{sec:experimental-comparison}
\begin{figure*}[htb!]
\centering
\includegraphics[width=0.98\textwidth]{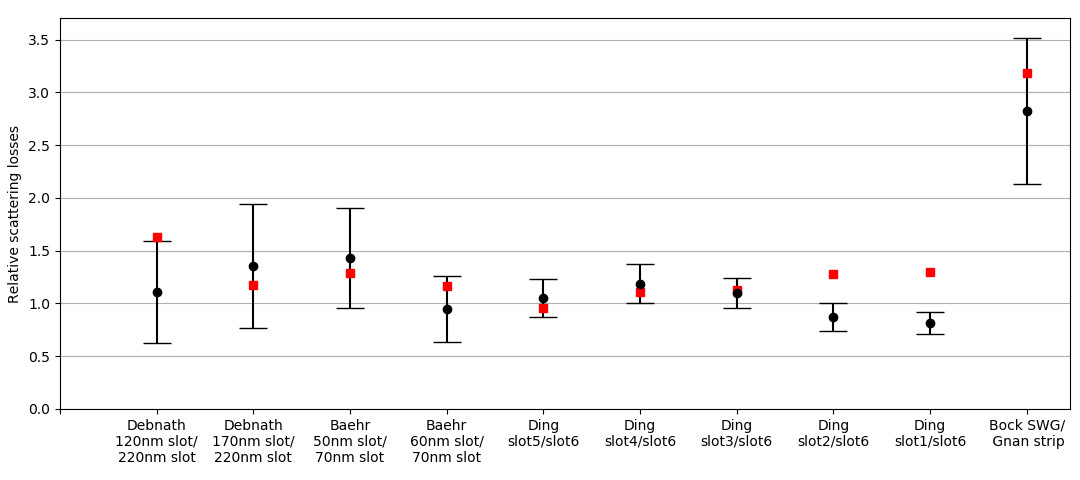}
\caption{Comparison of the ratio of scattering losses between different experimentally realized waveguide systems.  Red square markers denote loss ratios computed via the volume-current method described in this work and black circles denote reported loss ratios with associated error bars.  Slot waveguides reported by Debnath \emph{et al.} \cite{Debnath2016a} and Baehr-Jones \emph{et al.} \cite{Baehr-jones2005} are labelled by the slot size, while slot waveguides by Ding \emph{et al.} \cite{Ding2010} are labelled by ``slot1'', ``slot2'', etc. corresponding to different slot sizes and strip-loading values.  Bock \emph{et al.} \cite{Bock2010} and Gnan \emph{et al.} \cite{Gnan2008} report losses for SWG and strip waveguides fabricated via electron beam lithography.}\label{fig:comparison}
\end{figure*}

In order to demonstrate the utility and accuracy of this approach, we searched the literature for reports of fabricated strip, slot, and sub-wavelength grating silicon waveguides and their associated propagation losses \cite{Debnath2016a, Baehr-jones2005, Ding2010, Bock2010, Gnan2008}. We then applied the volume-current method to compute the relative scattering losses for each of 14 reported waveguide geometries. Since the waveguides of different geometries reported within each manuscript are fabricated using identical processing protocols which presumably result in identical or similar roughness characteristics, the reported losses are expected to accurately correspond to the different waveguide geometries.  The authors in Bock \emph{et al.} \cite{Bock2010} only report the loss for a single SWG structure and reference Gnan \emph{et al.} \cite{Gnan2008} to compare with strip waveguide losses. Both reported devices were fabricated using similar fabrication protocols (electron-beam lithography with hydrogen silsesquioxane resist and reactive-ion etching), and so we show in Fig. \ref{fig:comparison} the ratio of the reported propagation losses and our own numerically computed values using the reported waveguide geometries.
%
%
 
It is worth drawing attention to several features of Figure \ref{fig:comparison}. First and foremost, the experimental error bars associated with measured values of propagation loss are quite large, on the order of several dB/cm. Accurately comparing the propagation loss of different waveguides requires the fabrication of all considered geometries on one single substrate and in close proximity with each other (to avoid cross-wafer variations). The waveguide losses can be characterized through either the cut-back method (many waveguides of different lengths) or resonator devices \cite{Yariv2000}. The cut-back approach is susceptible to large variability in device-to-device coupling efficiency, whereas loss measurement using resonators only compute the loss for a small waveguide length \cite{Li2012}.  For these reasons, it is critical to measure a statistically significant number of devices. Often, information on the sample size and measurement errors are omitted in the literature.
 
Ding \emph{et al.} \cite{Ding2010} reported the striking result that two waveguide geometries (denoted by ``slot1'' and ``slot2'' in their work) have lower loss despite a larger modal overlap with the vertical sidewall interface. As a result of the higher modal overlap, our model predicts that ``slot1'' and ``slot2'' should actually perform worse. This inconsistency may originate from other experimental sources of loss (mask defects, scattering at the top surfaces of the partially etched silicon, absorption, etc.) that were modified by the change in waveguide geometry. Otherwise, the values reported in literature are consistent with values that our volume-current method predicts, which validates our approach as a fast, powerful technique for precisely comparing the relative performance of arbitrary waveguide systems \emph{before} investing time and resources towards fabricating real devices.

\section{Concluding Remarks}\label{sec:conclusion}

In this work, we numerically computed the relative performance enhancement of slot and SWG waveguides over traditional strip waveguides for on-chip sensing applications.  By explicitly computing the polarization statistics of randomly generated surface-roughness profiles (\href{link}{SI}), we confirmed that a simple flat shifted-boundary approximation is accurate for most commonly encountered surface roughness (correlation length greater than roughly 10 times the RMS roughness amplitude).  As a result, analytical formulas are available for the induced dipole moments from sidewall roughness.  For situations where the roughness correlation length is significantly shorter than the wavelength of light, the volume-current method allows the far-field and reflected radiation loss from this roughness to be determined with good accuracy using inexpensive numerical methods.  In particular, our approach benefits from significantly reduced simulation resolution requirements compared to brute-force simulation of waveguide structures with small perturbations, as the critical resolvable dimension is the waveguide geometry rather than the perturbation amplitude at the waveguide interface~\cite{Jaberansary2013}.  In addition, these brute-force techniques require many (or long) simulations to obtain statistically averaged effects of the randomly rough surfaces.  

Our approach can be readily extended to efficiently determine optimal waveguide geometries for other sensing modes, such as stimulated Raman spectroscopy~\cite{Zhao2018}, with a suitably defined sensing metric.  There are also a number of other material platforms to which this work can be extended, such as silicon nitride on silicon dioxide~\cite{Dhakal2013a, Dhakal2014, Holmstrom2016, Du2017}, titanium dioxide on silicon dioxide~\cite{Evans2016}, chalcogenide glass on insulator~\cite{Hu2009, Eggleton2011}, germanium or germanium-silicon on silicon~\cite{Chang2012,Liu2018}, silicon on sapphire~\cite{Baehr-Jones2010}, and more that are of great interest to the waveguide-integrated chemical sensing community. Because the waveguide geometries, materials, and figures of merit can change for different sensing processes and wavelengths, the conclusions about which geometries are better may also change. Lastly, there is interest in extending this work to quantify the sensing performance of additional waveguide geometries, such as photonic crystal waveguides~\cite{Meng2014} and horizontal slot waveguides~\cite{Sun2007}, to name just a few.



We believe this work and the methods presented will aid in the rational design of new waveguide geometries for photonic sensing applications.  Without figures of merit and efficient methods for computing the loss and sensing tradeoffs, it was difficult to predict whether waveguide geometries like slot and SWG waveguides will enhance sensing performance (due to increased field overlap with the sensing medium) or decrease sensing performance (from increased scattering losses).  With the techniques presented, it is possible to quantify the precise trade-off between these two competing factors for arbitrary waveguide geometries and material platforms.  This will drive future research in the area of on-chip sensing and other areas where propagation loss in waveguides plays an important role.
%
%
%
%
\setcounter{equation}{0}
\setcounter{figure}{0}
\setcounter{section}{0}
\renewcommand{\theequation}{S\arabic{equation}}
\renewcommand{\thefigure}{S\arabic{figure}}
\renewcommand{\thesection}{S\arabic{section}}
%
%
%
%

\vspace{1cm}
\section*{\LARGE{Supplementary information}}\refstepcounter{si}\label{s:SI}
The following sections provide supplementary information to ``Are slot and sub-wavelength grating waveguides better than strip waveguides for sensing?''.  First, we derive the relevant figures of merit (FOM) for absorption spectroscopy, refractometry, and waveguide-enhanced Raman spectroscopy.  Second, we outline methods for computing the polarizability of random Gaussian-correlated roughness.  Lastly, the exact volume-current methods for computing the relative scattering loss of strip, slot, and sub-wavelength grating (SWG) waveguides in the main text are presented in detail.

\section{Figure of merit for absorption spectroscopy and refractometry}

In the following, we explicitly compute the appropriate performance metrics for three different modes of sensing to illustrate their proportionality to the absorption spectroscopy and refractometry figure of merit, $\text{FOM}=\Gamma_\text{clad}/(\alpha_s \lambda)$.

\subsection{Absorption sensing}
On-chip absorption spectroscopic sensors measure the change in optical power when chemicals of interest in the waveguide cladding absorb light from the evanescent field. In this scheme, light of input power $P(z=0)$ ($z$ is the distance along the direction of propagation) at a known wavelength $\lambda$ is attenuated due to molecular absorption in the cladding region and scattering losses from waveguide imperfections. The rate of change in optical power $P$ in the presence of scattering losses (or other non-sensing losses, such as large Ohmic loss for plasmonic waveguides) $\alpha_s$ and absorption $\alpha_\text{abs}$ is:
\begin{equation}
\frac{dP}{dz} = -\Gamma_\text{clad} \alpha_\text{abs} P - \alpha_s P \label{eq:abs}
\end{equation}
where $\Gamma_\text{clad}$ is the external confinement factor that accounts for both slow-light effects and the fraction of electromagnetic energy residing in the waveguide cladding~\cite{Weber1974, Kumar1981, Mittra1986, She1990, Gupta1992, Themistos1995, Johnson2001a, gamma}
\begin{equation}
\Gamma_\text{clad} = \frac{n_\text{clad} c \epsilon_0 \int_\text{clad} |\vec{E}|^2 dx^2}{\int_\infty Re \{\vec{E}\times \vec{H}^*\} \hat{z} dx^2}
= \frac{n_g}{n_\text{clad}}\frac{\int_\text{clad} \epsilon |\vec{E}|^2 dx^2}{\int_\infty \epsilon |\vec{E}|^2 dx^2} \label{gamma}
\end{equation}
with $n_\text{clad}$ the cladding index, $n_g$ the group velocity, $c$ the speed of light, $\epsilon_0$ the vacuum premittivity, $\epsilon$ the material permittivity, and $E$ and $H$ the electric and magnetic fields of the waveguide modes. The absorption coefficient $\alpha_\text{abs} = \alpha_0 \cdot m$ is a wavelength-dependent attenuation coefficient that describes free-space optical absorption and depends on both the number density of molecules in the cladding, $m$, and the molecular absorption coefficient $\alpha_0$ (related to the transition dipole moment). The absorption coefficient tends to have strong peaks in the infrared due to vibrational and rotational transitions, so measuring power changes at wavelengths of high absorption is a sensitive and selective method for identifying specific gases or liquids~\cite{Lin2017}.  

The relevant quantity of interest is the change in optical power $P(L)-P(0)$ along a waveguide of length $L$ normalized by the input power $P(0)$ when a small amount of chemical/analyte is introduced to the sensing region.  The figure of merit (FOM) we care about is the corresponding sensitivity $S_\text{abs}=dp/dm$, where $p=(P(0)-P(L))/P(0)$ is the fractional optical power change.
\begin{equation}
S_\text{abs} = \frac{d}{dm} (1-e^{-\Gamma_\text{clad}\alpha_\text{abs}z - \alpha_s z}) = \Gamma_\text{clad}\alpha_\text{0}ze^{-\Gamma_\text{clad}\alpha_\text{abs}z - \alpha_s z}
\end{equation}
The sensitivity can then be maximized with respect to the waveguide length $z$:
\begin{align}
\frac{dS_\text{abs}}{dz} = 0 &= 1 -   z_\text{max} (\Gamma_\text{clad} \alpha_\text{abs} + \alpha_s)\\
z_\text{max} &= \frac{1}{\Gamma_\text{clad} \alpha_\text{abs} + \alpha_s}
\end{align}
The optimized sensitivity is then:
\begin{equation}
S_\text{max,abs} = \frac{\Gamma_\text{clad} \alpha_0}{\Gamma_\text{clad} \alpha_\text{abs} + \alpha_s} \cdot e^{-1}
\end{equation}
In most cases the fabrication-induced scattering losses is significantly larger than absorption (especially for sensing of trace molecules).  Thus, maximizing the external confinement factor and minimizing scattering losses is a suitable FOM since it maximizes the relative optical power change due to nearby absorbing chemicals.
\begin{equation}
S_\text{max,abs} = \frac{\Gamma_\text{clad} \alpha_0}{\alpha_s}e^{-1}\label{FOMabs}
\end{equation}

\subsection{Mach--Zehnder refractometry}

For a typical Mach--Zehnder refractometer-based sensor, which consists of a $1\times 2$ beam splitter, two waveguides with path lengths $z_1$ and $z_2$, and a $2\times 2$ (or $2 \times 1$) combiner, one measures a change in optical power at the output the interferometer to infer a change in the cladding refractive index $\Delta n_\text{clad}$.  As light propagates along one of the arms, it acquires an additional phase $\Delta \phi_1 = 2\pi \Gamma_\text{clad} \Delta n_\text{clad} z_1/\lambda + 2\pi n_g \Delta z/\lambda$ (where $\Delta z = z_2 - z_1$) due to the presence of chemicals in the cladding and the path length difference.  If we take $\Delta z = \lambda / (4n_g)$ (for convenience, and with no loss of generality) and $z_1 \sim z_2$ (which follows if $z \gg \Delta z$), the corresponding optical power at each of the two output ports is~\cite{fundamentals}:
\begin{align}
P_1 =& P(0) \cos^2(\pi \Gamma_\text{clad} \Delta n_\text{clad} z /\lambda + \pi/4) e^{-\alpha_s d}\\
P_2 =& P(0) \sin^2(\pi \Gamma_\text{clad} \Delta n_\text{clad} z /\lambda + \pi/4) e^{-\alpha_s d}
\end{align}
The sensitivity of our refractometer is the fractional power change ($p = (P_1 - P(0))/P(0)$) at one of the two ports for a small increase in analyte that produces an index shift $\Delta n_\text{clad}$ in the cladding:
\begin{equation}
S_\text{MZI} = \frac{dp}{d(\Delta n_\text{clad})} = \frac{z \Gamma_\text{clad}\pi}{\lambda}\cos\Big(\frac{2\pi \Gamma_\text{clad}\Delta n_\text{clad} z}{\lambda}\Big)e^{-\alpha_s z}
\end{equation}
The above metric is often called the figure of merit for Mach--Zehnder refractometers (the $e^{-\alpha_s z}$ term is often omitted, assuming no optical loss)~\cite{Heideman1990, Liu2013, Dullo2015, Bastos2018}.  For small index shifts, the cosine term tends towards 1 and the optimal arm length $z$ (that solves $dS/dz = 0$) is $z_\text{max}=1/\alpha_s$:
\begin{equation}
S_\text{max,MZI} = \frac{\Gamma_\text{clad} \pi}{\lambda \alpha_s} e^{-1}
\end{equation}

\subsection{Ring-resonator refractometry}

Likewise, ring-resonator-based refractometry sensors measure changes in the through-port power of a CW laser when molecules in the cladding change the index surrounding the ring waveguide core.  In this scheme, we assume a laser line is tuned to the steepest slope of the ring's resonance.  Corresponding changes in the index of the surroundings will shift the resonance and produce a corresponding change in output optical power.  The lineshape of a ring resonator nearby the resonance (assuming low cavity loss such that $\alpha d \ll 1$, with $d$ the ring roundtrip length) is given by~\cite{fundamentals}:
\begin{equation}
\frac{P_\text{output}}{P(0)} = 1 - \frac{1}{1 + \Big(\frac{2\pi (\nu-\nu_0) n_\text{eff}}{\alpha c_0}\Big)^2}
\end{equation}
where $\nu_0 = q \frac{c_0}{n_\text{eff}d}$ is the frequency of the $q$-th resonator mode, $c_0$ is the speed of light, and $\alpha$ is the resonator waveguide's loss per unit length.  In particular, the maximum slope occurs when the laser is tuned to $\nu = \nu_0 \pm \frac{\alpha c_0}{4\sqrt{3} n_\text{eff} \pi}$.  We can now define a sensitivity metric (similar to before) as being the fractional change in optical power $p = P_\text{output} / P(0)$ for a small change in the cladding index.  To first order, the entire resonance shifts by a frequency amount $\Delta \nu = -q \frac{c_0}{n_\text{eff}^2 d}\Gamma_\text{clad}\Delta n_\text{clad}$, so that the maximum sensitivity is:
\begin{equation}
S_\text{max,ring} = \frac{dp}{d(\Delta n_\text{clad})} = \frac{dp}{d\nu}\frac{d\nu}{d(\Delta n_\text{clad})} = \frac{3\sqrt{3} \pi q}{4 n_\text{eff} d} \frac{\Gamma_\text{clad}}{\alpha_s}
\end{equation}
where in the above we assume that the loss is predominantly due to scattering ($\alpha \approx \alpha_s$).

Here we showed that the sensitivity metric depends on the factor $\Gamma_\text{clad}/\alpha_s$ for optical intensity interrogation.  Furthermore, the sensitivity for other ring resonator sensing schemes such as wavelength interrogation are also proportional to $\Gamma_\text{clad}/\alpha_s$~\cite{Hu2009a}. In general, one can show that the sensitivity metric for almost all evanescent waveguide absorption sensors and refractometers includes the $\Gamma_\text{clad}/\alpha_s$ proportionality factor.  Other constants in the sensitivity metric account for the correct dimensionality and are in general not a function of the waveguide geometry.  Therefore, we choose the following dimensionless FOM as a suitable metric for comparing different waveguide geometries used in common refractometers and absorption sensors.
\begin{equation}
\text{FOM}_\Gamma = \frac{\Gamma_\text{clad}}{\alpha_s \lambda}
\end{equation}

\section{Spontaneous Raman spectroscopy figure of merit}
In waveguide integrated sensors that operate via spontaneous Raman spectroscopy, light at an initial wavelength $\lambda_p$ interacts with nearby molecules in the cladding and scatters light at a new wavelength $\lambda_n$.  The collection of this light serves as the signal of interest, and so the relevant figure of merit is the ratio of Raman scattered optical power collected by the waveguide $P_n$ to initial power $P_p$ in the waveguide, $P_n/P_p$.  Our semiclassical analysis (similar to derivations that use Fermi's golden rule~\cite{Dhakal2015, Jun2009}) accounts for the molecule's interaction with the electric field mediated by the full Raman polarizability matrix $\hat{\alpha}_\text{ram}$, and can be readily extended to the study of anisotropic scatterers and scattering mechanisms such as coherent Raman spectroscopy.  For a single molecular scatterer, the incident light excites a current source equal to the derivative of the change in dipole moment, $\vec{J}(\vec{x}) = -i\omega_n \Delta p(\vec{x},\omega_n)\delta(\vec{x}-\vec{x_0})=-i\omega_n \hat{\alpha}_\text{ram} \cdot \vec{E}_p(\vec{x}, \omega_p)\delta(\vec{x}-\vec{x_0})$, where $\omega_n$ is the new Raman scattered frequency, $E_p$ is the electric field of the incident waveguide mode (denoted by $p$) with frequency $\omega_p$, $\Delta p$ is the induced dipole moment, $\vec{x_0}$ is the location of the single molecule scatterer, and $\hat{\alpha}_\text{ram}$ is the spontaneous Raman polarizability of the molecule, a rank-2 tensor.  The total power radiated by a current source into each of the $i$ electromagnetic modes of the system (not waveguide modes) is given by~\cite{Oskooi2013}:
\begin{equation}
P = \frac{\pi}{4} \sum_{i} \frac{\Big|\int \vec{E}_{i}^*(\vec{x})\cdot \vec{J}(\vec{x}) dx^3\Big|^2}{\int_\infty \epsilon(\vec{x})|\vec{E}_{i}(\vec{x})|^2dx^3} \delta(\omega-\omega_{i})
\end{equation}
where the denominator serves to normalize the energy of the electric field.  For a constant cross-section waveguide (translationally invariant along the propagation direction $\hat{z}$), the total power coupled into the waveguide modes (indexed by $n'$) is:
\begin{equation}
P_\text{wg} = \frac{\pi}{4} \sum_{n'} \int dk_z \frac{\Big|\int_\text{cs} \vec{E}_{n'}^*(\vec{x},k) \cdot \vec{J}(\vec{x})\Big|^2}{\int_\text{cs} \epsilon(\vec{x})|\vec{E}_{n'}(\vec{x},k)|^2} \delta(\omega-\omega_{n'}(k_z))\label{eq:MDOS}
\end{equation}
where the ``cs'' subscript denotes integration over the entire two-dimensional cross-section, $\vec{x}$ vectors refer to the 2-dimensional position vector, and $k_z$ is the wave vector in the $\hat{z}$ direction (the integrand is sometimes referred to as the ``mutual density of states'', or MDOS~\cite{Mcphedran2004}).  We now wish to compute the power radiated into a single waveguide mode (denoted by $n$), and so the above is simplified to:
\begin{equation}
P_{n} = \frac{\pi}{4} \frac{\Big|\int_\text{cs} \vec{E}_{n}^*(\vec{x},\omega_{n})\cdot \vec{J}(x) \Big|^2}{\int_\text{cs} \epsilon(\vec{x})|\vec{E}_{n}(\vec{x}, \omega_{n})|^2} \rho_{1D}(\omega_{n})
\end{equation}
In the above, $\rho_{1D} = 1/(\pi v_g)$ is the 1-dimensional density of states for the waveguide arising from the delta function with a Jacobian factor~\cite{Arfken2000}.  Direct substitution of the driven current source expression for a single-molecule scatterer yields:
\begin{equation}
P_{n} = \frac{\omega_{n}^2}{4 v_g(\omega_{n})} \frac{\Big|\vec{E}_{n}^*(\vec{x_0},\omega_{n})\cdot \hat{\alpha}_\text{ram}\cdot \vec{E}_p(\vec{x_0}, \omega_p)\Big|^2}{\int_\text{cs} \epsilon(\vec{x})|\vec{E}_{n}(\vec{x},\omega_{n})|^2}
\end{equation}
We typically only care about the amount of power scattered into the waveguide for a given number density of scatterers $m$ (number per unit volume) in the cladding region over an infinitesimal distance $dz$ along the propagation direction.  Therefore we take an average of the collected power over all inequivalent positions in the cladding (denoted by the subscript ``clad''):
\begin{equation}
\langle P_{n} \rangle = \Big(\frac{\omega_{n}^2 \cdot m \cdot dz}{4 v_g(\omega_{n})}\Big)  \frac{\int_\text{clad} \Big|\vec{E}_{n}^*(\vec{x},\omega_{n})\cdot \hat{\alpha}_\text{ram}\cdot \vec{E}_p(\vec{x}, \omega_p)\Big|^2}{\int_\text{cs} \epsilon(\vec{x})|\vec{E}_{n}(\vec{x},\omega_{n})|^2}
\end{equation}
For simplicity, we assume the polarizability $\hat{\alpha}_\text{ram}$ is independent of position, but in general this could change if the density varies. Finally, we normalize the average Raman-scattered power to the input laser power $P_p$:
\begin{align}
\frac{\langle P_{n} \rangle}{P_p} =&
\Big(\frac{\omega_{n}^2 \cdot m \cdot dz}{4 v_g(\omega_{n})v_g(\omega_p) }\Big) \times \\
&\frac{\int_\text{clad} \Big|\vec{E}_{n}^*(\vec{x},\omega_{n})\cdot \hat{\alpha}_\text{ram}\cdot \vec{E}_p(\vec{x}, \omega_p)\Big|^2}{ (\int_\text{cs} \epsilon(\vec{x})|\vec{E}_p(\vec{x},\omega_p)|^2)(\int_\text{cs} \epsilon(\vec{x})|\vec{E}_{n}(\vec{x},\omega_{n})|^2)}
\end{align}
If we assume the molecules are isotropic and uniformly distributed, then the polarizability matrix is simply a scalar and we can factor it from the integral in the numerator.  When comparing waveguide geometries for the purpose of maximizing the signal of interest for waveguide Raman spectroscopy, the relevant design parameters are the two factors of group velocity $v_g$ that account for slow light effects, the electric field strength of the incident waveguide mode at $\omega_p$, and the electric field strength of the new waveguide mode at $\omega_{n}$.  Thus, the relevant quantity of interest is the Raman gain coefficient $\beta$ in units of inverse distance:

\begin{align}
\beta& = \frac{\langle P_n \rangle}{P_p}  \frac{1}{dz} =\\
&\Big(\frac{\alpha_\text{ram}^2 \cdot \omega_n^2 \cdot m}{4 v_g(\omega_n) v_g(\omega_p)}\Big) \frac{\int_\text{clad} \Big|\vec{E}_n^*(\vec{x},\omega_n) \vec{E}_p(\vec{x}, \omega_p)\Big|^2}{(\int_\text{cs} \epsilon(\vec{x})|\vec{E}_p(\vec{x},\omega_p)|^2)(\int_\text{cs} \epsilon(\vec{x})|\vec{E}_n(\vec{x},\omega_n)|^2)}
\end{align}

For waveguides with period $\Lambda$, Eq.~\ref{eq:MDOS} can be equivalently rewritten in terms of 3-dimensional integrals over the entire unit cell:
\begin{equation}
P_\text{wg} = \frac{\pi}{4} \sum_{n'} \frac{\Lambda}{2\pi} \int_{-\frac{\pi}{\Lambda}}^{\frac{\pi}{\Lambda}} dk_z \frac{\Big|\int_\text{unit cell} \vec{E}_{n'}^*(\vec{x},k) \cdot \vec{J}(\vec{x})\Big|^2}{\int_\text{unit cell} \epsilon(\vec{x})|\vec{E}_{n'}(\vec{x},k)|^2} \delta(\omega-\omega_{n'}(k_z))
\end{equation}
and we note that the delta function in $\vec{J}(\vec{x})$ above is 3-dimensional in contrast to the 2-dimensional delta function used in Eq.~\ref{eq:MDOS} (thus preserving units).  If we proceed in the same manner as before, we arrive at the Raman gain coefficient for periodic structures:
\begin{align}
\beta_\text{SWG} &=
\Big(\frac{\alpha_\text{ram}^2 \cdot \omega_n^2 \cdot m}{4 v_g(\omega_n) v_g(\omega_p)}\Big) 
\Big(\frac{\Lambda}{2\pi} \Big) \times \\
&\frac{\int_\text{clad} \Big|\vec{E}_n^*(\vec{x},\omega_n) \vec{E}_p(\vec{x}, \omega_p)\Big|^2}{(\int_\text{unit cell} \epsilon(\vec{x})|\vec{E}_p(\vec{x},\omega_p)|^2)(\int_\text{unit cell} \epsilon(\vec{x})|\vec{E}_n(\vec{x},\omega_n)|^2)}
\end{align}

\begin{figure*}[h!]
\centering
\fbox{\includegraphics[width=\textwidth]{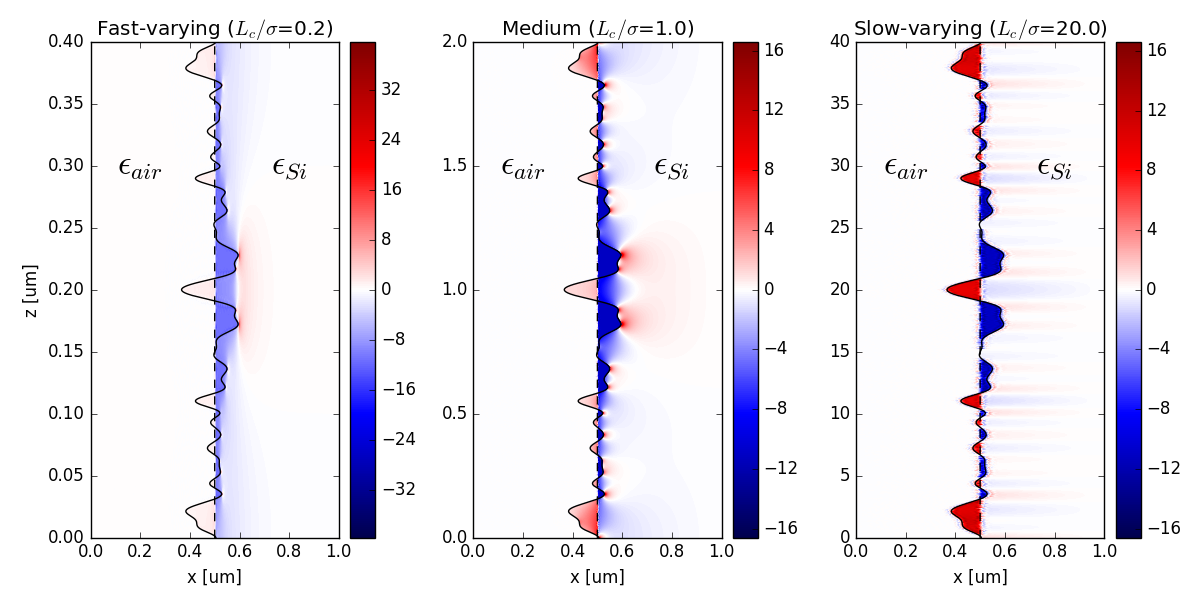}}
\caption{Computed changes in the perturbed polarization density $\Delta P(\vec{r}) \cdot \hat{z}$ along a rough waveguide sidewall with short ($L_c/\sigma=0.2$), medium ($L_c/\sigma=1.0$), and long ($L_c/\sigma=20$) correlation lengths with respect to the RMS roughness $\sigma$.  Solid black line shows the perturbed sidewall, and dashed black line shows the straight, unperturbed sidewall.}\label{fig:surface}
\end{figure*}

Now, we may readily incorporate the effects of scattering loss due to waveguide imperfections.  This is typically described by the scattering loss attenuation coefficient $\alpha$ with units of inverse length.  The change in our signal $P_n$ per unit length is similar to Eq.~\ref{eq:abs}, but with an additional loss term to account for the Raman gain coefficient and propagation losses at both signal- and Raman-scattered wavelengths:
\begin{equation}
\frac{dP_n}{dz} = -\alpha_n P_n + \beta P_p e^{-\alpha_p z}
\end{equation}
where $\alpha_p$ and $\alpha_n$ denote the scattering loss at the incident wavelength and new Raman scattered wavelength, respectively.  Solving this differential equation yields:
\begin{equation}
\frac{P_n(z)}{P_p(0)} = \frac{\beta e^{-\alpha_n z}}{\alpha_n - \alpha_p}[e^{z(\alpha_n-\alpha_p)}-1]\\
\end{equation}
Further simplification can be made by assuming the scattering losses at incident and signal wavelengths are equal.
\begin{equation}
\lim_{\alpha_n \rightarrow \alpha_p} P_n(z) = \beta P_p z e^{-\alpha_{n,p} z} \label{eq:sig}
\end{equation}
Our goal is then to maximize the signal power by optimizing $z$, the length of the waveguide.  We can see from Eq.~\ref{eq:sig} that too short of a waveguide has insufficient length for the light to interact with the analyte and generate a signal, while a waveguide that is too long loses a significant amount of incident and signal light from scattering losses.
\begin{align}
\frac{dP_n(z)}{dz} = 0 &= \beta P_p e^{-\alpha_{n,p} z_\text{max}} - \alpha_{n,p} z_\text{max} \beta P_p e^{-\alpha_{n,p} z_\text{max}}\\
z_\text{max} &= 1/\alpha_{n,p}
\end{align}
If we consider a waveguide integrated Raman sensor with optimized waveguide length, the relevant figure of merit for sensing is then simply the signal power at $z=z_\text{max}$ divided by the input power at $z=0$.
\begin{equation}
\frac{P_n(z_\text{max})}{P_p(0)} = \frac{\beta}{\alpha_{n,p}} \cdot e^{-1}
\end{equation}
In this work, we are interested in determining only the optimal waveguide geometries for spontaneous Raman sensing.  Therefore, constants such as the Raman polarizability and the number density of scatterers are not relevant and are set to unity in our comparison. In most practical Raman spectrometers, $\omega_n \approx \omega_p$. This is in part due to increased Raman scattering efficiencies at higher optical frequencies such as $\bar{\nu}$ = 12,500 - 20,000 cm$^{-1}$ ($\lambda_p = 500 - 800$nm), since $\alpha_\text{ram}$ scales with $\omega^4$~\cite{Ferraro2002}.  Meanwhile, typical Raman shifts of interest correspond to low-energy vibrational or rotational transitions in the region of 500-1,500 cm$^{-1}$.  Therefore, our dimensionless figure of merit that involves only terms relevant for comparing different waveguide geometries (i.e. constants and prefactors set equal to 1) is:
\begin{equation}
\text{FOM}_{\beta} = \frac{\beta}{\alpha_s} \propto \frac{1}{\alpha_s} \cdot \frac{n_g^2(\omega_p)}{n_\text{clad}^2} \frac{\int_\text{clad}|\vec{E}(x_0, \omega_p)|^4 dx_0}{(\int_\text{cs} \epsilon(x)|\vec{E}(x,\omega_p)|^2 dx)^2} \label{FOMram}
\end{equation}
and for periodic structures,
\begin{equation}
\text{FOM}_{\beta_\text{SWG}} = \frac{\beta_\text{SWG}}{\alpha_s} \propto \frac{\Lambda}{\alpha_s} \cdot \frac{n_g^2(\omega_p)}{n_\text{clad}^2} \frac{\int_\text{clad}|\vec{E}(x_0, \omega_p)|^4 dx_0}{(\int_\text{unit cell} \epsilon(x)|\vec{E}(x,\omega_p)|^2 dx)^2} \label{FOMramswg}
\end{equation}

\begin{figure*}[h!]
\centering
\fbox{\includegraphics[width=\textwidth]{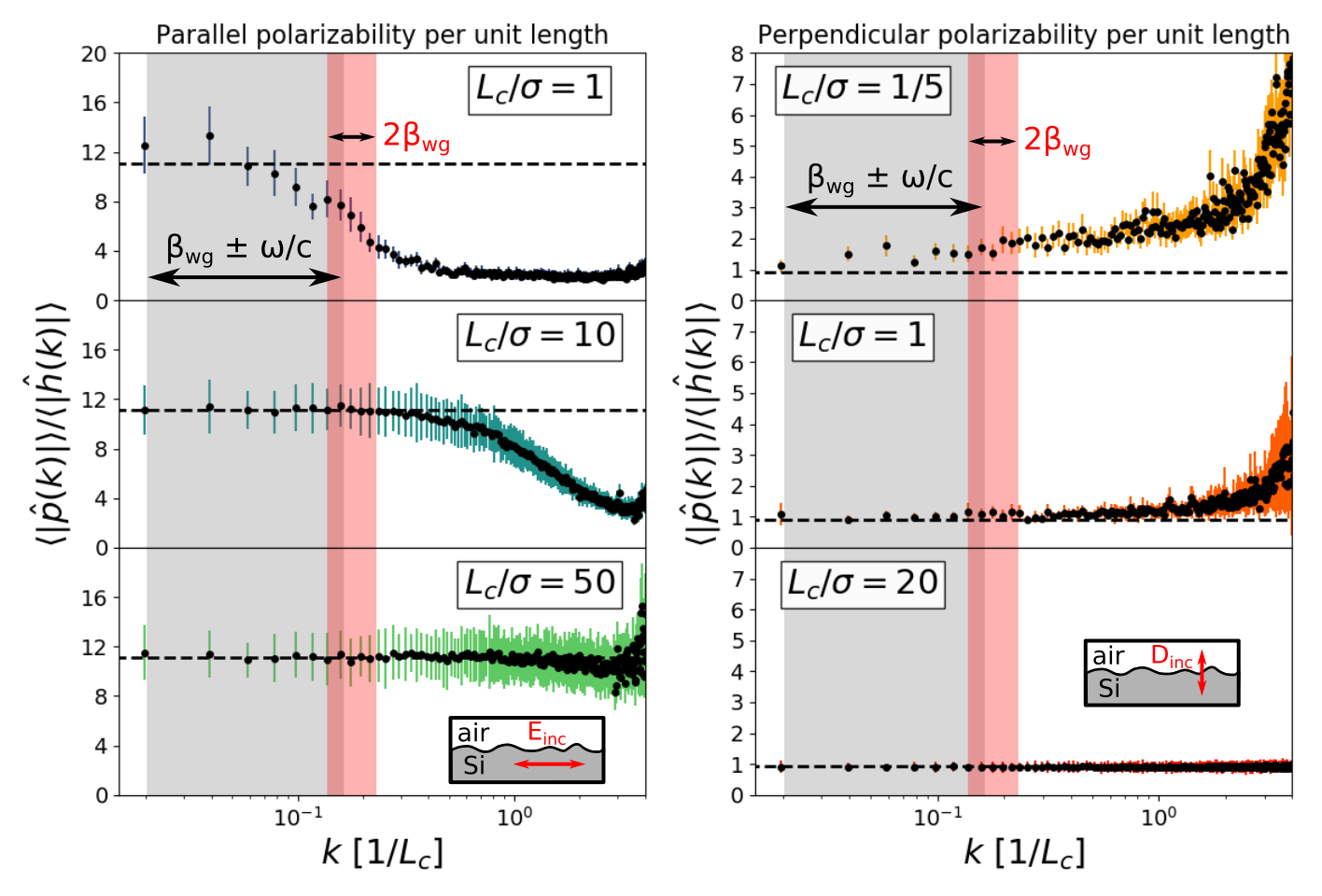}}
\caption{Numerically computed Fourier components of the polarizability for a randomly generated surface with Gaussian correlation length and a unit-valued incident field parallel (left) and perpendicular (right) to the interface for different ratios of $L_c/\sigma$.  Black dashed lines correspond to the analytical polarizabilities of $\alpha_\parallel=11.1$ and $\gamma_\perp = 0.92$ for a flat shifted silicon/air interface.  Gray shaded regions correspond to Fourier-components of the roughness that scatter light into the far-field and red shaded regions correspond to components that reflect light in the backwards direction ($\beta_\text{wg}$ corresponds to the range of propagation coefficients for all waveguide-geometries considered in this text, and is scaled assuming a roughness correlation length of 75\,nm and wavelength of $\lambda$=1550\,nm).  Colored errorbars denote the standard error of the mean for multiple computations.}\label{fig:FTpol}
\end{figure*}

\section{Polarizability of rough surfaces}
In the volume-current method, it is critically important to understand the nature of the perturbed dipole moments from perturbations at a dielectric interface.  For small, localized perturbations (such as bumps or point-defects), the electric polarizability $\alpha_\parallel$ and $\gamma_\perp$ relates the magnitude of the current source $\vec{J}$ to the incident field.  However, for roughness that exists along the entire interface, our goal is to understand the statistical properties of the ensemble of dipole moments in the quasi-static limit.  In the following, we computationally verify that the flat shifted-interface polarizability is an accurate approximation for Gaussian-correlated randomly rough sidewalls so long as the correlation length $L_c$ is considerably larger than the root-mean-square (RMS) amplitude of the perturbation height $\sigma$.  

\begin{figure*}[tb!]
\centering
\fbox{
\includegraphics[width=\linewidth]{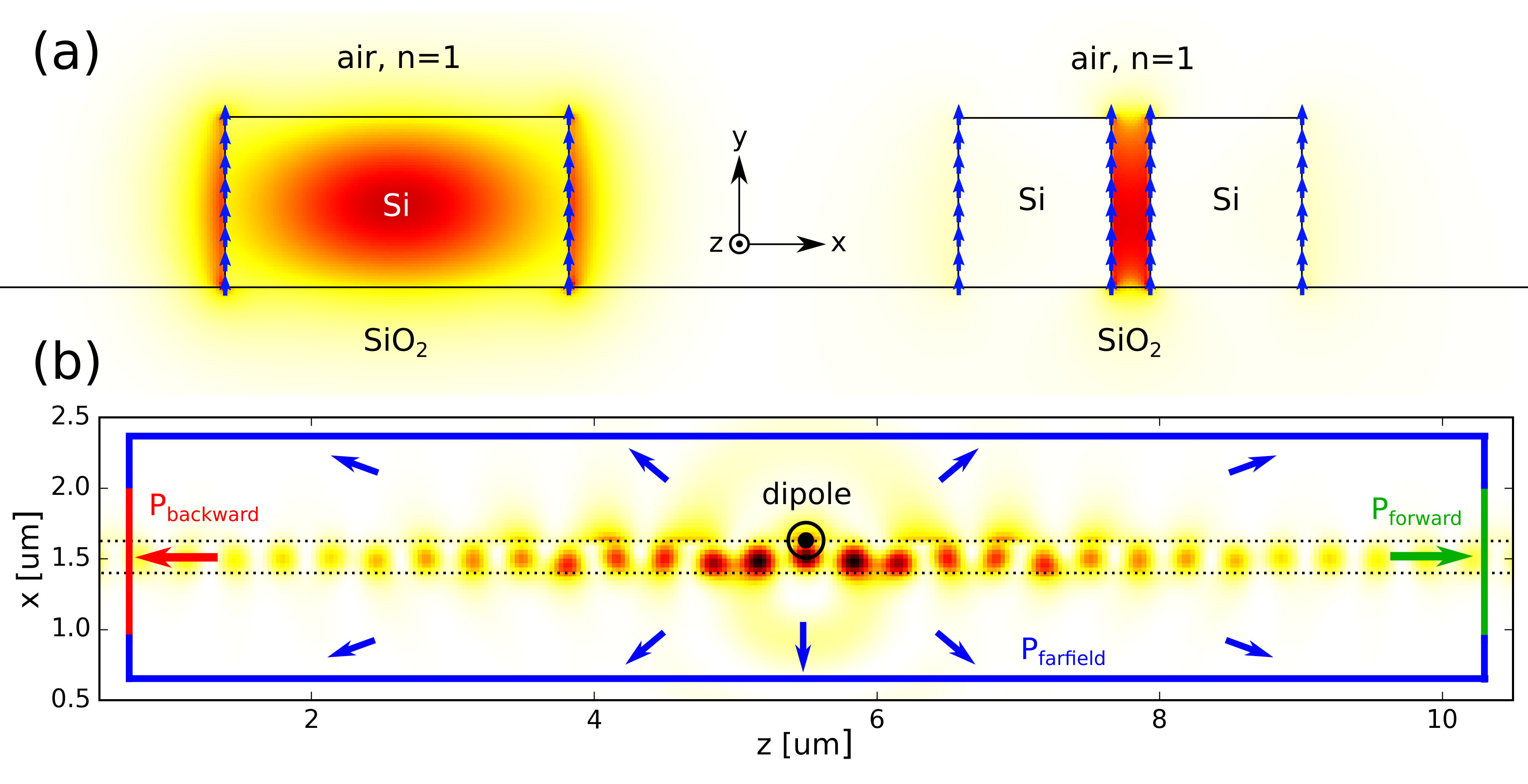}}
\caption{(a) Cross-sectional image of the electric field magnitude $|\vec{E}|^2$ for a dipole radiating along the surface of a strip waveguide. The blue solid line indicates the position of power flux monitors meausuring the amount of radiation into the far-field. The green and red solid lines denote locations of power flux monitors that measure the radiation that couples back into the forward and backwards propagating waveguide mode. Simulation resolution shown above and in all calculations is 32 pixels/$\mu$m. (b) Cross-sectional image in the $\hat{x}-\hat{y}$ plane for both strip (left) and slot (right) structure waveguides with $|\vec{E}|^2$ shown in red. For each FDTD simulation, a line of current sources are placed at each inequivalent sidewall position (only one for each strip waveguide geometry and two for each slot waveguide geometry), with the appropriate complex amplitude retrieved from the waveguide mode's electric field strength at each point.}\label{fig:fdtd_dipoles}
\end{figure*}

To determine the appropriate limits for which the flat-shifted interface is valid, we determined the Fourier components of the 1-dimensional electrostatic polarizability density (along the interface-direction) by numerically computing the induced dipole moment density (per unit length) for randomly generated surfaces. Zero-mean sidewall functions $h(z)$ with a Gaussian autocorrelation function $R(z') = \sigma^2 \exp(-z'^2/L_c^2)$ were generated to represent the roughness profile, where $\sigma$ is the root mean square (RMS) roughness amplitude and $L_c$ denotes the roughness correlation length~\cite{Oppenheim1999}. This was done by filtering Gaussian white noise with the appropriate filter functions specified by the desired roughness autocorrelation function~\cite{Billah1990, Kasin1995, Young2000, Lu2005}. We then solved the corresponding electrostatic Poisson equation using FEniCS, an open-source partial differential equation solver~\cite{Alnaes2015}, to obtain the resulting charge distribution for silicon/air interfaces with a sidewall profile $h(z)$, as shown in Figure~\ref{fig:surface}. For perpendicular incident fields with constant field $D_\perp$, we use periodic boundary conditions at the edges of the interface and Neumann boundary conditions for simulation boundaries parallel to the interface. For parallel incident fields with constant field $E_\parallel$, we specify the corresponding electrostatic potential via appropriate Dirichlet boundary conditions along the simulation boundaries. Each simulation window used was $320\cdot L_c$ long and $40 \cdot \sigma$ wide, with an adaptive finite-element mesh. The dipole moment density corresponding to each specific roughness profile $h(z)$ was then computed from the spatial charge distribution through~\cite{Jackson1999}:
\begin{align}
p_{1D,\perp}(z) &= \frac{1}{\Delta z}\int_{-\infty}^{\infty} \int_{z-\Delta z/2}^{z+\Delta z/2} (x-x_0)\cdot \Delta \rho(\vec{x}) d\vec{x}\label{eq:linear_dipole_moment}\\
p_{1D,\parallel}(z) &= \frac{1}{\Delta z}\int_{-\infty}^{\infty} \int_{z-\Delta z/2}^{z+\Delta z/2} (\Delta \vec{P}(\vec{x})\cdot \hat{z}) d\vec{x}
\label{eq:linear_dipole_moment2}\end{align}
where $\Delta z$ is the pixel size along the interface, $\Delta \vec{P}$ is the change in electrostatic polarization density ($\vec{P} = \epsilon_0(\epsilon - 1) \vec{E}$) and $\hat{x}$ is normal to the interface (perpendicular to $\hat{z}$).  In principle, we can use the same expression for both perpendicular and parallel components, but we found the above expressions converged faster at a given finite resolution. The dipole moment density per unit perturbation height $h(z)$ is then related to the polarizability through $p_\parallel/h = \alpha_\parallel E_\parallel$ (and likewise $p_{\perp}/h = \gamma_\perp D_\perp$).  Using this technique, we also performed a set of validation tests on flat-shifted boundaries and cylindrical/square perturbations to ensure that our method (and mesh resolution) accurately reproduces the correct electrostatic polarizabilities.

Using this technique to compute the induced dipole moment densities, we investigated the effect of different surface roughness statistics, quantified by the dimensionless ratio $L_c/\sigma$.  Figure~\ref{fig:FTpol} shows the ratio of the averaged Fourier components (in spatial-frequency space) of the dipole moment density $\hat{p}(k)$ to the Fourier components of the surface profile $\hat{h}(k)$.  For a locally flat interface, the induced dipole moments are proportional to the perturbed height, and so far from $L_c$ we expect $\hat{p} \sim \hat{h}$.  In the limit of long correlation lengths (large $L_c/\sigma$) our computations recover the expected polarizability of a flat-shifted interface, shown by dashed black lines in Figure~\ref{fig:FTpol}. In general, the surface roughness will act as a diffraction grating~\cite{Lacey1990} that couples incident light of wavevector $\beta$ to $k' = k_z + \beta$ with a strength proportional to the roughness induced dipole moment $\hat{p}(k_z)$.  However, light can only be scattered into the light cone or reflected back into the waveguide, which limits the relevant frequency components to being within $k' = \pm \frac{\omega}{c}$ (far-field radiation) and at $k' = 2\beta$ (reflection).  The appropriate ranges for all waveguide geometries considered in the main text are depicted by the shaded regions in Figure~\ref{fig:FTpol} (assuming a roughness correlation length of $L_c$=75\,nm and $\lambda$=1550\,nm).

As $L_c\rightarrow \infty$, it must be true that the locally flat result is recovered.  In our analysis, we observe the useful result that for even relatively short correlation lengths on the order of $L_c = 10\sigma$, the Fourier components of the numerically computed polarizability match the simple shifted-boundary approximation by $\leq 1.2$\,\% for the interface-parallel components and $\leq 0.5$\,\% for the interface-perpendicular components. In state-of-the-art SOI waveguides, the correlation length of line-edge roughness was experimentally assessed to be between 50 to 100~nm, whereas the root mean square (RMS) roughness is typically between 0.5 to 2~nm~\cite{Xia2007, Wood2014,Lee2015,Wang2014}, corresponding to $L_c/\sigma$ ratios on the order of 25--200. Therefore, we may accurately model the roughness-induced effective current-sources as flat-shifted interfaces for a large number of practical silicon photonic waveguide devices.

\section{Volume-current method calculations}

The relative scattering losses for each waveguide structure was determined by calculating the relative powers per unit volume of sidewall roughness (defined as the volume of the ``bumps'' on a surface) scattered into both the far-field $P_\text{ff}$ and reflected backwards in the waveguide $P_\text{r}$. We restricted our focus to vertically symmetric ($y$-direction symmetric) line-edge roughness, and therefore modeled the roughness with 1-dimensional line current sources (in all 3-directions) with corresponding local amplitudes and phases proportional to the modal electric field, as depicted in Figure~\ref{fig:fdtd_dipoles}(a).  The electric and displacement field of each waveguide mode was computed using the MIT Photonic Bands software package~\cite{Johnson2001}, and special care was taken to ensure that the normalization of the 3-dimensional SWG modes are equivalent to the normalization of the 2-dimensional strip and slot waveguide modes.  We also make sure to normalize the input modal fields by \emph{power}, rather than \emph{energy}, to properly account for group velocity effects, and since our metric of interest is the scattered \emph{power} per unit input power $P_s/P_\text{in}$.  The current-source is determined by:
\begin{equation}
\vec{J} = -i\omega \Delta h (\Delta \epsilon E_\parallel - \epsilon \Delta (\epsilon^{-1})D_\perp)\delta(\vec{x})
\label{eq:currentstrength}
\end{equation}
For each waveguide geometry and each inequivalent sidewall position, the corresponding line of current sources is then simulated using MEEP, a freely available finite-difference time domain (FDTD) software package~\cite{Oskooi2010}.  The amount of power scattered into the far-field, versus forward and backward scattered power are determined by monitoring the power flux (at a wavelength of 1550\,nm) through the simulation boundaries, as shown in Figure~\ref{fig:fdtd_dipoles}(b).  Each FDTD simulation is computed with a resolution of 32 pixels/$\mu$m and a domain size of 10\,$\mu$m\,$\times$\,2.5\,$\mu$m\,$\times$\,2.5\,$\mu$m with perfectly matched layers (PML) at the boundaries, except for SWGs where we place adiabatic absorbers at the $z_\text{min}$ and $z_\text{max}$ boundaries.  The total power loss $P_s$ (the sum of the far-field power $P_\text{ff}$ and the reflected power $P_{r}$) is then averaged over all sidewall positions and computed per unit waveguide length.

\section*{Acknowledgments}
We would like to thank Alexander McCauley for originating the polarization-statistics approach for roughness with short correlation lengths.

This work is supported by the National Science Foundation under Award No. 1709212 and in part by the Army Research Office under contract number W911NF-13-D-0001.

\bibliographystyle{ieeetr}
\bibliography{Edge_roughness}


\end{document}